\documentclass[12pt]{amsart}

\newcommand{\ed}{

\end{document}
}

\newcommand{\x}{\times}
\newcommand{\xor}{\oplus}
\newcommand{\mx}[1]{\left(\begin{matrix}#1\end{matrix}\right)}

\newtheorem{thm}{Theorem}
\newtheorem{lem}[thm]{Lemma}
\theoremstyle{definition}
\newtheorem{defn}[thm]{Definition}
\theoremstyle{remark}
\newtheorem{rem}[thm]{Remark}

\author{Boaz Tsaban}
\thanks{Supported by the Koshland Center for Basic Research.}
\email{boaz.tsaban@weizmann.ac.il}
\address{Department of Mathematics,
Weizmann Institute of Science, Rehovot 76100, Israel}

\title[Cryptanalysis of TF-1]{\textsf{NOTE}\\[0.5cm]Theoretical cryptanalysis of the Klimov-Shamir number generator TF-1}

\begin{document}
\begin{abstract}
The internal state of the Klimov-Shamir number generator TF-1 consists of
four words of size $w$ bits each,
whereas its intended strength is $2^{2w}$.
We exploit an asymmetry in its output function to show that
the internal state can be recovered after having $2^w$ outputs,
using $2^{1.5w}$ operations. For $w=32$ the attack is practical,
but for their recommended $w=64$ it is only of theoretical interest.
\end{abstract}

\maketitle

\section{Generalized TF-1 generators}

The \emph{Klimov-Shamir number generator TF-1} was introduced in \cite{KSHand}
and is based on the methods developed in \cite{KS04} and references therein.
This is an iterative pseudorandom number generator.
Its internal state consists of four words $a,b,c,d$, of size $w$ bits each.
$C_1,C_2,C_3,C$ are fixed constants chosen to optimize several properties (which
are not relevant for our analysis).
The update function of the generator is defined as follows.\footnote{In the following description,
$\land,\lor,\xor$ denote bitwise logical \emph{and}, \emph{or}, and \emph{xor}, respectively, and
addition and multiplication are always carried modulo $2^w$.}
$$\mx{a\\ b\\ c\\ d}\mapsto
\mx{
a & \xor & s & \xor & 2c\cdot (b\lor C_1)\\
b & \xor & s\land a & \xor & 2c\cdot (d\lor C_3)\\
c & \xor & s\land a\land b & \xor & 2a\cdot (d\lor C_3)\\
d & \xor & s\land a\land b\land c & \xor & 2a\cdot (b\lor C_1)
}$$
where
$$s = (C+(a\land b\land c\land d))\xor (a\land b\land c\land d).$$
After each update, an output value
$$S(a+c)\cdot (S(b+d)\lor 1)$$
is extracted, where $S$ is the function swapping the upper and lower halves of its input,
i.e., $S(x)=x/2^{w/2} + x\cdot 2^{w/2}$ for each $x=0,\dots,2^w-1$
where ``$/$'' denotes integer division.

Earlier variants of this generator were cryptanalyzed in several works,
see for example \cite{MiSa, BeRe}.
None of the earlier attacks applies to the present generator, though,
since the present output function is more complicated.
We will present an attack on a generalized family of TF-1
generators, containing the Klimov-Shamir generator as a particular
case.

\begin{defn}[Klimov-Shamir \cite{KS04}]
$T:\{0,1\}^{m\x w}\to\{0,1\}^{n\x w}$ is a
\emph{T-function} if,
for each $k=1,\dots,w$,
the first $k$ columns of $T(X)$ depend only on the first $k$ columns of $X$.
\end{defn}

Note that, using the convention that words from $\{0,1\}^w$ are
written such that the leftmost bit is the least significant one,
the update function of a TF-1 generator is a T-function.

Following is a generalization of the family of TF-1 generators.
The fact that we pose no restriction on its function $F$ (and still
are able to cryptanalyze it as shown below) seems to be of special interest.

\begin{defn}
A \emph{generalized TF-1} generator
consists of an update function $T_1:\{0,1\}^{4\x w}\to \{0,1\}^{4\x w}$
and output auxiliary functions $T_2,F: \{0,1\}^{4\x w}\to \{0,1\}^w$.
$T_1$ and $T_2$ are T-functions,
but $F$ can be any efficiently computable function.
Its internal state is a matrix $A\in\{0,1\}^{4\x w}$,
The update function is
$$A\mapsto T_1(A).$$
After each update, an output value
$$S(T_2(A))\cdot (F(A)\lor 1)$$
is extracted.
\end{defn}

\section{Cryptanalysis}

Generators with poor statistical properties are
not suitable for cryptographic usage. We therefore restrict
attention to the nondegenerate cases.

\begin{lem}\label{enumerate}
Assume that $T: \{0,1\}^{4\x w}\to \{0,1\}^w$ is a (mildly) random-looking T-function,
$k,l\in\{1,\dots,w\}$, and $l\le k$.
If the first $l-1$ columns of $X$ are known and $T(X)=0$,
then the list of all possibilities for columns $l,\dots,k$ of $X$
can be enumerated in (roughly) $2^{3(k-l)}$ operations.
\end{lem}
\begin{proof}
First check all $2^4$ possibilities for the $l$th column of $X$.
Only about $2^3$ should give $0$ at the $l$th bit of $T(A)$.
For each of them, check all $2^4$ possibilities for the $l+1$th bit.
Again about $2^3$ of which will survive. Continue in this manner.
The total number of operations is roughly
$$2^4+2^3\cdot 2^4+(2^3)^2\cdot 2^4 + \dots + (2^3)^{k-l-1}\cdot 2^4 \approx 2\cdot 2^{3(k-l)}.$$
Note that there is no need to store the resulting tree in memory, since
the search in the tree could be of ``depth first'' type, i.e., follow each
branch up to its end before moving to the next branch.
\end{proof}

\begin{rem}
For the function $T((a,b,c,d)^t)=a+c$ used in TF-1,
the enumeration as in Lemma \ref{enumerate} is trivial:
Just enumerate $(a,b,-a,d)^t$ where $a,b,d\in\{0,1\}^k$.
Note further that $0$ plays no special role in the proof of Lemma \ref{enumerate}
and it can be replaced by any constant.
\end{rem}

\begin{thm}\label{attack}
Assume that $G$ is a generalized TF-1 generator which is (mildly) random-looking.
Then the internal state of $G$ can be recovered from roughly $2^w$ output words,
using roughly $2^{1.5w}$ operations.
\end{thm}
\begin{proof}
Scan the output sequence until an output word $0$ is found (this requires
roughly $2^w$ output words). Denote the internal state at this point by $A$.
Then
$$S(T_2(A))\cdot (F(A)\lor 1) = 0.$$
As $F(A)\lor 1$ is relatively prime to $2^w$,
we have that $S(T_2(A))=0$, and therefore $T_2(A)=0$.

Use Lemma \ref{enumerate} with $l=1$ and $k=w/2+1$ to enumerate
the $2^{3k}$ possibilities for the first $k$ columns of $A$.
During the enumeration, compute for each possibility
the first $k$ columns of $A'=T_1(A)$ and of $T_2(A')$.
The $k$th bit of $T_2(A')$ should be equal to
the least significant bit of the next output word.
This rules out about half of the suggested solutions.
Checking about one more step will rule out about half of the remaining solutions, etc.
Algorithmically, continue updating and checking until a contradiction
is found (or until a solution survives more than $3k$ steps)
and then move to the next suggested solution. On average
this requires two steps per suggested solution.

Having completed the above $2^{3k+1}$ operations, the first $k$ columns of $A$ are known.
Use Lemma \ref{enumerate} again to go over all possibilities for columns $k+1,\dots,w$ of $A$.
Now there are only $2^{3k-6}$ possibilities, and each of them gives a complete
knowledge of the internal state and can thus be checked by computation of
one or two output words. The total amount of operations is roughly
$$2^{3k+1}+2^{3k-6}\approx 2^{3k+1}=2^{1.5w+4}=16\cdot 2^{1.5w}.\qedhere$$
\end{proof}

\section{Examples}

Any generalized TF-1 generator for words of $32$ bits has an internal state
of size $128$ bits and intended strength $2^{64}$. By Theorem \ref{attack},
the whole internal state can be recovered
from $2^{32}$ output words (i.e., $16$ gigabytes) using
$16\cdot 2^{1.5\cdot 32}=2^{52}$ operations.
These parameters are practical.

Any generalized TF-1 generator for words of $64$ bits has an internal state
of size $256$ bits and intended strength $2^{128}$. By Theorem \ref{attack},
the internal state can be recovered
from $2^{64}$ output words using $16\cdot 2^{1.5\cdot 64}=2^{100}$ operations.
In this setting, our attack is only of theoretical interest.

\subsection*{Acknowledgments}
We thank Alexander Klimov and the referees for their comments.

\ed